\documentclass{article}
\usepackage{spconf,amsmath,graphicx}
\usepackage{enumitem}
\usepackage[hidelinks]{hyperref}
\usepackage{physics} 

\ninept

\title{Towards zero-shot Text-based voice editing using acoustic context conditioning, utterance embeddings, and reference encoders}
\name{Jason Fong$^1$, Yun Wang$^2$, Prabhav Agrawal$^2$, Vimal Manohar$^2$, Jilong Wu$^2$, Thilo Köhler$^2$, Qing He$^2$ \thanks{$^1$Work performed during internship.}}
\address{$^1$University of Edinburgh, $^2$Meta AI}
\begin{document}
%
\maketitle
\begin{abstract}
Text-based voice editing (TBVE) uses synthetic output from text-to-speech (TTS) systems to replace words in an original recording. Recent work has used neural models to produce edited speech that is similar to the original speech in terms of clarity, speaker identity, and prosody. However, one limitation of prior work is the usage of finetuning to optimise performance: this requires further model training on data from the target speaker, which is a costly process that may incorporate potentially sensitive data into server-side models. In contrast, this work focuses on the \textit{zero-shot} approach which avoids finetuning altogether, and instead uses pretrained speaker verification embeddings together with a jointly trained reference encoder to encode utterance-level information that helps capture aspects such as speaker identity and prosody. Subjective listening tests find that both utterance embeddings and a reference encoder improve the continuity of speaker identity and prosody between the edited synthetic speech and unedited original recording in the zero-shot setting. 
\end{abstract}
\begin{keywords}
Speech synthesis, text-based speech editing
\end{keywords}

\begin{figure*} 
  \centering
  \includegraphics[width=\linewidth]{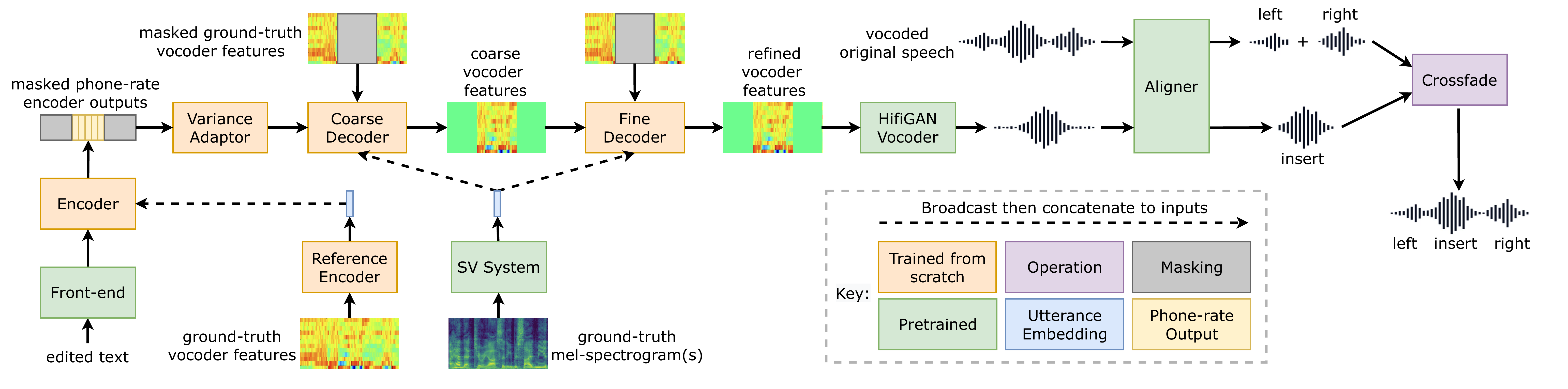}
    \vspace*{-7mm}
  \caption{Overview of our Text-based voice editing system performing inference time editing of some original speech.}
  \label{fig:overview}
  \vspace{-3mm}
\end{figure*}

\section{Introduction}
\label{sec:intro}

In recent years there has been rapid improvement in text-to-speech (TTS) technology due to the widespread adoption of neural network based acoustic models \cite{wang2017tacotron,arik2017deep} and vocoders \cite{oord2016wavenet,kalchbrenner2018efficient,kong2020hifi}; Subsequently, it is now \textit{possible} to convincingly use synthetic speech to replace some content in place, as it can potentially match the original speech in terms of clarity, speaker similarity, and prosody. We refer to this line of research as text-based voice editing (TBVE). 


TBVE however is a more challenging task than standard TTS, since it must situate synthetic speech within natural speech in an imperceptible way, there must be a  high level of consistency between the edited and unedited regions. Although prior TBVE approaches have achieved promising results \cite{wang2022campnet,tae2021editts,tan2021editspeech,morrison2021context}, they are still far from being \textit{completely} natural. Furthermore,  TBVE models often adopt a one-shot or few-shot paradigm where they are finetuned on target speaker data in order to improve speaker similarity, however finetuning is not applicable in some scenarios as it is a costly process that bakes voice data into server models, which may have privacy implications.

Considering the privacy-based limitations of prior work, we solely focus on the challenging task of increasing TBVE performance in the completely \textit{zero-shot} setting. That is, our models are not finetuned on any data from target speakers. We only use speech information extracted from the single recording to be edited, and do not incorporate any extracted information back into the model. The zero-shot scenario is challenging as extracting disentangled speaker and utterance characteristics from original audio is not a solved problem. In this work we structure our models to tackle this challenge. The main contributions of this paper are as follows: 
\begin{itemize}[topsep=3pt, leftmargin=*]
  \setlength{\itemsep}{1pt}
  \setlength{\parskip}{0pt}
  \setlength{\parsep}{0pt}
    \item We demonstrate that TBVE works in our Fastspeech 2 \cite{ren2020fastspeech} inspired TTS framework which uses a variance adaptor to predict phone-wise frame durations rather than an attention-based encoder-decoder approach \cite{wang2017tacotron,wang2022campnet}. 
    \item We demonstrate that even without the usage of speaker embeddings, speaker information can be extracted effectively from acoustic context.
    \item We demonstrate that utterance embeddings and a reference encoder improve continuity in the zero-shot setting, potentially removing the need for finetuning on extra data from target speakers.
\end{itemize}

\section{Related work}
\label{sec:related}

\subsection{Speaker Adaptation for Neural TTS}

Neural TTS models can be trained on the data of a single speaker and produce high quality speech \cite{wang2017tacotron}. To provide multispeaker TTS functionality, one can then train separate single-speaker TTS models \textit{from scratch} on the data of additional speakers. 

However, an often more desirable approach to multispeaker TTS is to \textit{adapt} a previously trained model to new speakers. One method named finetuning involves taking a pretrained TTS model, and then training it further on a target speaker, which is often done with less data and for fewer training iterations \cite{arik2018neural, chen2018sample, chen2021adaspeech, choi2020attentron}. Although this saves the compute cost of training multiple target speaker specific models from scratch, one downside is that one finetuned model must be stored for \textit{each} target speaker. This is often infeasible due to storage constraints and the difficulty of maintaining many models. 

A second method of speaker adaptation performs zero-shot adaptation via speaker embeddings \cite{jia2018transfer, cooper2020zero}. We use this approach in our TBVE model. This involves training a TTS system to reconstruct the speech of multiple speakers, with the help of embeddings obtained from a pretrained speaker verification system. At test time, speaker embeddings can be extracted from target speakers not in the training corpus in order to perform zero-shot speaker adaptation. The downside of this approach however is that the speaker characteristics often are not an exact match for any given target speaker since fewer model parameters are adapted than finetuning. In this work we counteract this weakness by adopting a reference encoder that is trained jointly with the TBVE task and as such can  capture details within an utterance not captured by the  speaker verification model that are beneficial for voice cloning and therefore also voice editing.

\subsection{Text-based voice editing}

Prior neural TBVE systems attempt to generate synthetic speech that matches the unedited original recording, by conditioning a generative model on some new words to be inserted and some acoustic context extracted from original speech \cite{wang2022campnet,tae2021editts,tan2021editspeech,morrison2021context}. One such example is CampNet \cite{wang2022campnet} from which we draw inspiration from for two of our methods. Firstly we use the same `Context-Aware Mask Prediction' training scheme which attempts to mirror the test time editing process: during training a random sequence of frames are masked out from the original speech's spectrogram, the resulting masked spectrogram is then concatenated to the decoder's inputs, which is trained to reconstruct the entire spectrogram. Secondly we adopt the same `coarse-fine' two stage decoder where a coarse decoder predicts vocoder features which are then refined by a fine decoder which also predicts the same vocoder features. Our model however diverges from CampNet as we use  FastSpeech 2's variance adaptor to model durations, rather than an attention-based sequence-to-sequence approach. Furthermore, our model does not finetune on any target speaker data, which is required by CampNet to get optimal results. In contrast, our models uses utterance embeddings and a reference encoder to help maintain acoustic continuity.

\section{Text-based voice editing model}
\label{sec:modeldescription}

Our TBVE editing model is visualised in Figure \ref{fig:overview}. At its core is a non-autoregressive FastSpeech 2 inspired architecture. The multi-layer transformer encoder consumes phone-rate feature vectors generated by the front-end from raw text to produce encodings. The variance adaptor then consumes these encodings to predict the $f_0$ and duration for each phone timestep. Predicted durations are then used to upsample the concatenated text encodings and predicted $f_0$'s up to the frame-rate. These features are then fed to the decoder. 

We implement a two-stage decoding process in order to produce refined acoustic predictions. The coarse and fine decoders, both of which are feed-forward transformer encoders, predict in parallel frame-rate vocoder features consisting of a 1-dim $f_0$ vector, a 13-dim MFCC vector, and a 5-dim periodicity vector that are used to condition a neural vocoder. The fine decoder receives as input the vocoder features predicted from the coarse decoder.

In addition to the core TTS model, we propose several enhancements in order to incorporate speaker and contextual information from the unedited region with the goal of improving the continuity of the edit with respect to speaker similarity and prosody in the zero-shot setting. Firstly, we use speaker or utterance embeddings. These are extracted from a pretrained speaker verification model. The embeddings are broadcasted and then concatenated to both the coarse and fine decoder inputs. 

Secondly, we also concatenate the output of a reference encoder \cite{skerry2018towards}, which is used to summarise utterance level characteristics from the original unedited audio. Previously these have been adopted primarily for prosody transfer. In this work we use them to capture utterance level prosodic \textit{or} speaker variations that are not captured by the pretrained speaker verification system. We also include a variational version of the reference encoder \cite{hsu2018hierarchical, liu2021vara} as it may potentially capture a better disentangled representation that exhibits finer control at inference time.  

Thirdly, to incorporate fine-grained acoustic context \cite{chen2021adaspeech} we additionally concatenate to the decoder inputs vocoder features (MFCCs, $f_0$, periodicity) where frames corresponding to the replaced words in the original speech are masked out with zeros \cite{wang2022campnet}. Word to frame alignments are obtained using a pretrained alignment model. Additionally we concatenate to the encoder inputs a two-value 32-dim mask token to further emphasise to the model which frames correspond to the masked region. 

To train the model we apply an L2 loss over the frames that correspond to the masked out phones between predicted and ground-truth vocoder features. This L2 loss is applied to both the coarse and fine decoder predictions, with a weighting of 1.0 and 10.0 respectively in the overall loss function. During training we randomly alter the mask rate from 0.2 to 1.0. A mask rate of 0.2 means that a sequence that is equal in length to 20\% of the phones is masked out, and the model must predict the frames corresponding to these masked phone positions. A mask rate of 1.0 means that no acoustic context is supplied and the model must predict all frame timesteps, which is equivalent to standard TTS. 

\section{Experiments}
\label{sec:experiments}

In this section, we detail our choices regarding data, model configurations, and evaluation in order to test the following hypotheses:

\begin{enumerate}[leftmargin=1.0cm, nolistsep, noitemsep,topsep=2pt,parsep=0pt,partopsep=0pt]
    \item[\textbf{H\textsubscript{1}:}] \textbf{Does concatenating masked contextual acoustic features improve continuity?}
    \item[\textbf{H\textsubscript{2}:}] \textbf{Does usage of speaker or utterance embeddings improve continuity?}
    \item[\textbf{H\textsubscript{3}:}] \textbf{Does usage of acoustic reference encodings improve continuity?}
    \item[\textbf{H\textsubscript{4}:}] \textbf{Does the fine decoder improve continuity?}
\end{enumerate}

\subsection{Data}
\label{subsec:data}

We trained our TBVE model and HifiGAN vocoder \cite{kong2020hifi} on an internal dataset of 378 hours of clean speech data from 95 speakers with a variety of US English accents. We evaluated our systems on 4 unseen speakers (2 male and 2 female). Using a pretrained speaker verification system, we extracted 32-dimensional utterance embeddings from each utterance, and calculated speaker embeddings by averaging all the utterance embeddings for a given speaker.

\begin{table}
\centering
  \caption{Gender-partitioned speaker similarity results from the first listening test. Most conditions had a 90\% confidence interval of ± 0.07/0.08 MOS score for female speakers and ± 0.10/0.11 for male speakers}
  \vspace{1mm}
  \label{tbl:qf1_results_table_male_female}
  \includegraphics[width=0.80\linewidth]{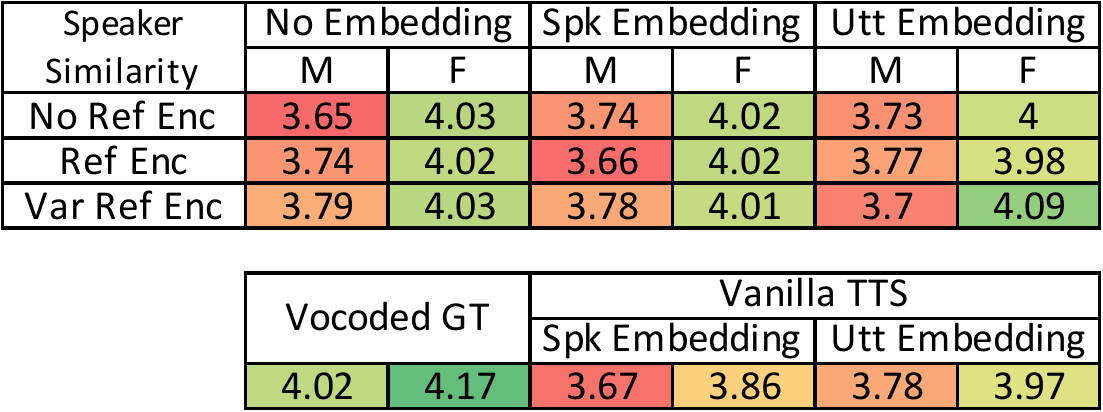}
  \vspace{-3mm}
\end{table}

\subsection{Model architecture details}
\label{subsec:modelarch}

Our TBVE model uses the same encoder, variance adaptor, and decoder architectures as Fastspeech 2 \cite{ren2020fastspeech}. The encoder and decoders all consist of 4 feed-forward transformer blocks which transform the input features into a hidden sequence. The variance adaptor, which predicts variance information (durations, $f_0$) for each phone timestep, consists of 2 one-dimensional convolutional layers with ReLU activations each followed by layer normalisation and dropout. 

Our reference encoder architecture, which takes inspiration from \cite{skerry2018towards, jia2018transfer}, consumes an utterance of frame-rate vocoder features and outputs a single timestep vector summary. It consists of 3 sequential one-dimensional convolutional layers that downsample the vocoder features in time; each layer is followed by batch normalisation and ReLU activation and has a kernel size of 3, stride of 2, padding of 1, and a hidden dimension of 128. The third convolutional layer is followed by a single layer bi-directional gated recurrent unit (GRU) which generates a 64-dimensional summary of its inputs in both the forward and backward directions by taking the final hidden output of both recurrent units. These forward and backward summaries are then concatenated together and passed through a final output linear layer to form a 32-dimensional reference encoding. 

The variational form of the reference encoder \cite{zhang2019learning} instead passes the GRU's output through two separate linear layers, one that predicts gaussian means and the other predicting gaussian variances, which are then reparameterised  to a 32-dimensional latent z \cite{kingma2013auto} which is then passed to a final output layer generating a 32-dimension reference encoding. The latent z is trained to match a gaussian distributed variable by using a KL divergence loss term in the overall loss function. The weight given to this loss term during training is 0.001. 

\subsection{Model configurations for listening test}
\label{subsec:modelconfigs}


For our main study we generate samples from 12 conditions. The topline condition is unedited ground truth audio that has been converted to vocoder features then vocoded back to the waveform domain. We vocode the ground truth for a fair comparison since all other conditions involve vocoding. The two baseline conditions  are Vanilla TTS models (one uses speaker embeddings and the other utterance embeddings) that predict acoustic features for the entire edited sentence. Target edited words are then extracted from the waveform using forced alignment, and fused with the vocoded ground truth by crossfading. The final 9 conditions are our proposed TBVE systems trained on all combinations of \{no embedding, speaker embedding, utterance embedding\} and \{no reference encoder, reference encoder, variational reference encoder\}. At inference time these models only generate the target edited words, which are then extracted by forced alignment and then fused with the vocoded ground truth. 

For our secondary study, we trained additional models to help answer \textbf{H\textsubscript{1}} and \textbf{H\textsubscript{4}}. These models are summarised in Table \ref{tbl:qf3_results_table}, where `Proposed model' uses utterance embeddings, a non-variational reference encoder, and concatenated masked acoustic context.

\subsection{Evaluation design}
\label{subsec:evaluation}

In our first listening test, we examined three aspects of the synthetic speech in the edited region by asking participants whether clarity, speaker identity and prosody have high continuity between the edited and unedited region. For each generated sample we provided the utterance and the transcript with the \textit{two} edited words highlighted, and asked participants to rate on a scale of 1-5 the continuity between the edited and unedited region for each aspect. We generated 60 stimuli from each condition using 60 held out test set sentences (evenly distributed over 4 held out speakers, two female, and two male) for a total of 720 stimuli. We recruited 425 participants for this test. For our second listening test, we presented only audios and asked participants ``Has this been edited?''. To create a total of 440 samples we use the vocoded ground truth of 220 unique utterances, and an edited version of each one, evenly distributed over the other 11 conditions. We recruited 60 participants for this test. Finally for our secondary study, we perform a third listening test, where we follow the same methodology as our first listening test, but only generate stimuli from the 6 conditions in Table \ref{tbl:qf3_results_table}, creating 360 stimuli in total. We recruited 175 participants for this test.


\begin{table}
  \caption{Results from the first listening test (female speakers only). Raters graded each stimuli on a scale of 1-5 in terms of how continuous three aspects (Clarity, Speaker Similarity, Prosody) were between the edited words and the original unedited speech. Each system had a 90\% confidence interval of ± 0.08 or 0.09 MOS score. Abbreviations are mapped as follows - 
  Ref: Reference, Enc: Encoder, Spk: Speaker, Utt: Utterance, GT: Ground-Truth, Clar: Clarity, Pros: Prosody.}
  \vspace{1mm}
  \label{tbl:qf1_results_table}
  \centering
  \includegraphics[width=0.85\linewidth]{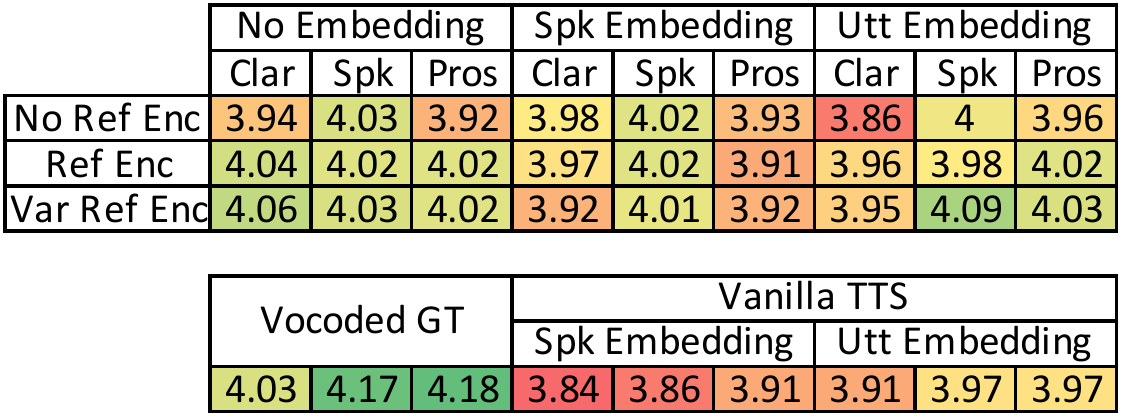}
  \vspace{-3mm}
\end{table}


\section{Results}
\label{sec:results}

\subsection{Performance disparity between male and female speakers.} Upon listening to the listening test stimuli, it was evident that speaker similarity was significantly better for the two female speakers, with synthetic speech from the male speakers being very dissimilar to the unedited speech in terms of speaker similarity. Gender-partitioned speaker similarity results from our first listening test support this observation and are shown in Table \ref{tbl:qf1_results_table_male_female}. We found that our proposed TBVE model, when editing female speech, is at the least on par with the unedited vocoded ground-truth of the male speakers. Subsequently, for the rest of this section we chose to present  results from only the female speakers as the poor  male speaker speaker similarity make cross-condition comparisons less clear. To investigate the cause of this disparity we examined t-SNE  speaker embedding plots, but did not find any anomalies due to gender. In future work we wish to further investigate and resolve this disparity.

\begin{figure}
  \centering
\includegraphics[width=1.0\linewidth]{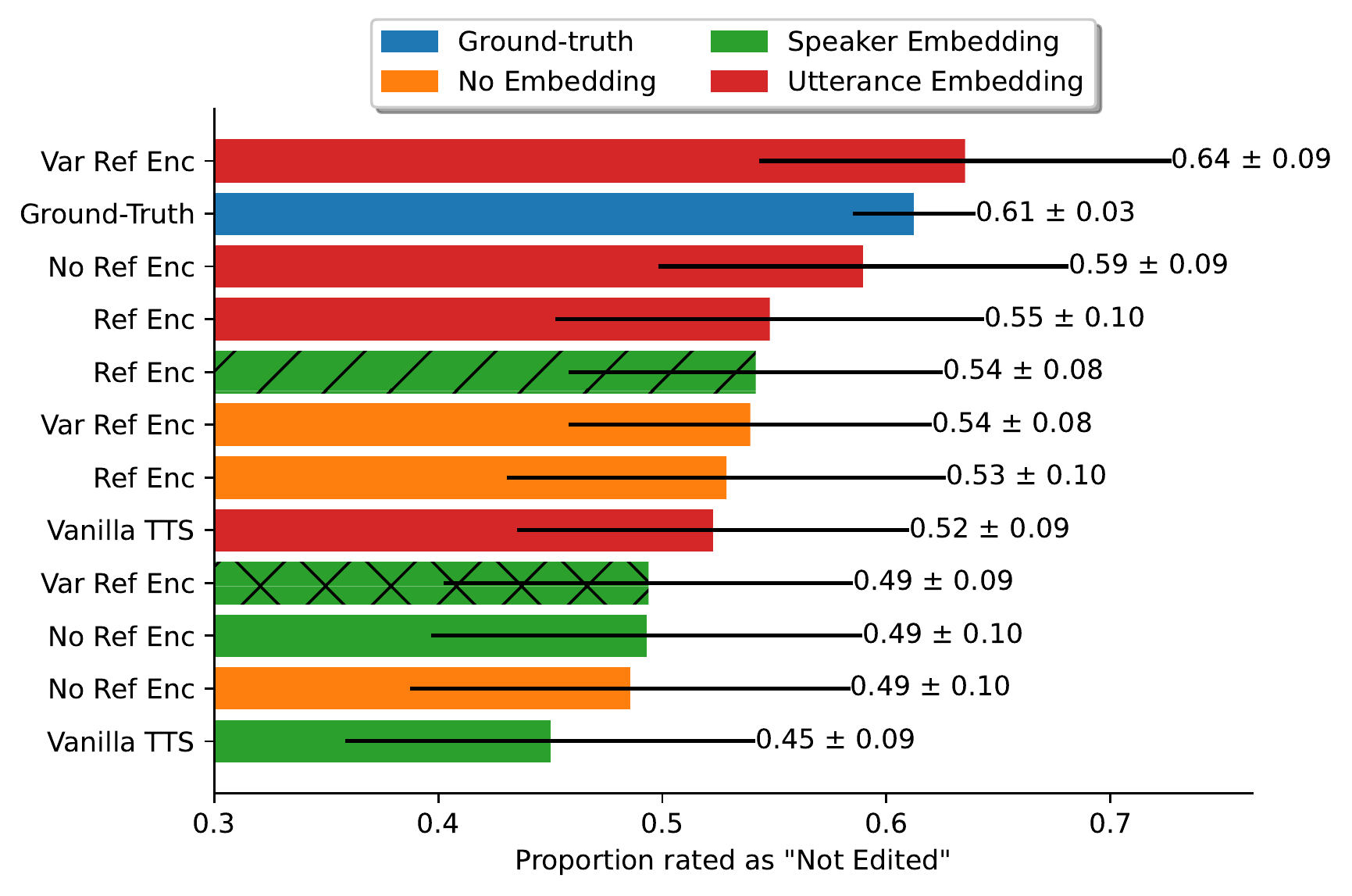}
\vspace*{-7mm}
  \caption{Results from the second listening test with 90\% confidence intervals (female speakers only). Raters judged whether each stimuli had been edited or not. No hatching indicates no reference encoder, diagonal hatching indicates use of a reference encoder, and cross hatching indicates use of a variational reference encoder.}
  \label{fig:qf2_results}
  \vspace{-3mm}
\end{figure}

\subsection{H\textsubscript{1}: Does concatenating masked contextual acoustic features improve continuity?} We identify three pieces of evidence that suggest that contextual acoustic features benefit our TBVE model. Firstly, in Table \ref{tbl:qf1_results_table} we find that our basic proposed system (`No Embedding', `No Ref Enc') outperforms the Vanilla TTS with speaker embeddings system in terms of speaker similarity, suggesting that our model can successfully extract speaker identity information from the acoustic features. Furthermore in Figure \ref{fig:qf2_results} we found that our basic proposed system is  rated as ``Not Edited'' with a higher likelihood than Vanilla TTS.

Secondly, in Table \ref{tbl:qf1_results_table} we find that our proposed model with speaker embeddings but no reference encoder is better than Vanilla TTS with speaker embeddings across all continuity aspects. This improvement is also exhibited in Figure \ref{fig:qf2_results}.
Since the same speaker embeddings are supplied to both models this suggests that our proposed model can capture valuable utterance-level characteristics from the acoustic features. 

Finally in Table \ref{tbl:qf3_results_table}, we find that `Proposed model' has improved continuity in all aspects over `- Acoustic Context', this suggests that the finer grain detail contained within the acoustic features are additionally beneficial for improving continuity. This is perhaps important at the two edit boundaries where a model's predictions should smoothly transition to and from the unedited original recorded speech.

\subsection{H\textsubscript{2}: Does usage of speaker or utterance embeddings improve continuity?} We see in Table \ref{tbl:qf1_results_table} and Figure \ref{fig:qf2_results} one clear trend: using utterance embeddings improves speaker and prosody continuity over speaker embeddings. However we find that when a reference encoder is used,  speaker embeddings  are actually worse than  no embeddings, especially in terms of clarity and prosody continuity. This suggests that the reference encoder is capturing   utterance-specific features which  model prosody better than speaker embeddings. This might be because the reference encoder is trained jointly from scratch to optimise the TBVE loss whereas the speaker verification model is frozen after pretraining. 

Finally we also observe that utterance embeddings compared with no embeddings are worse in terms of clarity continuity  when a reference encoder is also used, this  suggests that the utterance embeddings and the reference encoder capture similar types of utterance level information that are entangled together in a way that does not generalise well at test time.

\begin{table}
\centering
  \caption{Results from the third listening test (female speakers only). Each system had a 90\% confidence interval of ± 0.09 or 0.10 MOS score}
  \vspace{1mm}
  \label{tbl:qf3_results_table}
  \includegraphics[width=0.57\linewidth]{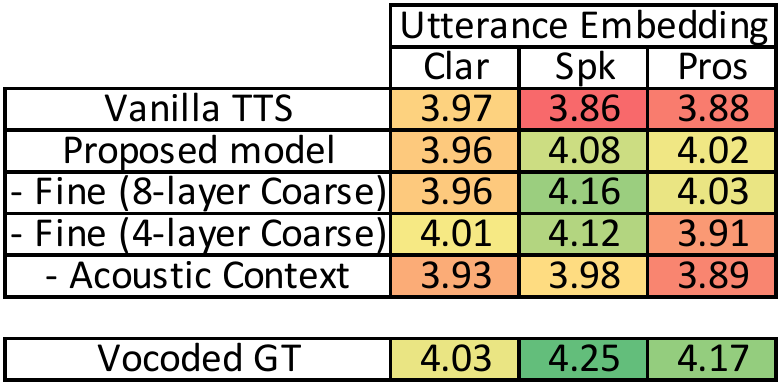}
  \vspace{-3mm}
\end{table}

\subsection{H\textsubscript{3}: Does usage of acoustic reference encodings improve continuity?} From Table \ref{tbl:qf1_results_table} and Figure \ref{fig:qf2_results} we find that the reference encoder is broadly beneficial for clarity and prosody continuity for the systems that either do not have embeddings or use utterance embeddings. Additionally it seems that the standard reference encoder largely performs similarly to the variational version. However we found that using the variational reference encoder along with utterance embeddings  had the strongest speaker similarity among the proposed systems in Table \ref{tbl:qf1_results_table} \textit{and} even outperformed the ground-truth condition in Figure \ref{fig:qf2_results}. This condition performs  very strongly and also suggests that high speaker similarity is crucial in improving perception of whether a stimuli has been edited or not.

\subsection{H\textsubscript{4}: Does the fine decoder improve continuity?} In Table \ref{tbl:qf3_results_table} we see that the proposed model is outperformed in terms of clarity and speaker continuity when the fine decoder is removed. The 4 and 8-layer coarse decoder only models overall perform better, with the 8-layer outperforming in terms of speaker and  prosody continuity. This suggests that it is beneficial to have a deeper decoder rather than opting for a coarse-fine two-stage decoder setup when speaker and prosody continuity are paramount.

\section{Conclusion}
\label{sec:conclusion}

In this work we have introduced our text-based voice editing system that uses three sources of information to improve the continuity between edited and the original speech in terms of speaker identity and prosody. Our evaluation shows that (1) concatenating masked acoustic context features is as effective as using speaker embeddings from a pretrained speaker verification system, (2) using utterance embeddings is superior to using speaker embeddings, and (3) utterance-level encodings from a jointly trained acoustic reference encoder performs well even without the usage of speaker or utterance embeddings. This work has demonstrated that improving zero-shot voice editing is possible with only model-based innovations. Future work should seek to improve the extraction of utterance-level information from original speech to further improve continuity performance.


\clearpage

\bibliographystyle{IEEEbib}
\bibliography{refs}

\begin{thebibliography}{10}

\bibitem{wang2017tacotron}
Yuxuan Wang, RJ~Skerry-Ryan, Daisy Stanton, Yonghui Wu, Ron~J Weiss, Navdeep
  Jaitly, Zongheng Yang, Ying Xiao, Zhifeng Chen, Samy Bengio, et~al.,
\newblock ``Tacotron: Towards end-to-end speech synthesis,''
\newblock {\em arXiv preprint arXiv:1703.10135}, 2017.

\bibitem{arik2017deep}
Sercan~{\"O} Ar{\i}k, Mike Chrzanowski, Adam Coates, Gregory Diamos, Andrew
  Gibiansky, Yongguo Kang, Xian Li, John Miller, Andrew Ng, Jonathan Raiman,
  et~al.,
\newblock ``Deep voice: Real-time neural text-to-speech,''
\newblock in {\em International Conference on Machine Learning}. PMLR, 2017,
  pp. 195--204.

\bibitem{oord2016wavenet}
Aaron van~den Oord, Sander Dieleman, Heiga Zen, Karen Simonyan, Oriol Vinyals,
  Alex Graves, Nal Kalchbrenner, Andrew Senior, and Koray Kavukcuoglu,
\newblock ``Wavenet: A generative model for raw audio,''
\newblock {\em arXiv preprint arXiv:1609.03499}, 2016.

\bibitem{kalchbrenner2018efficient}
Nal Kalchbrenner, Erich Elsen, Karen Simonyan, Seb Noury, Norman Casagrande,
  Edward Lockhart, Florian Stimberg, Aaron Oord, Sander Dieleman, and Koray
  Kavukcuoglu,
\newblock ``Efficient neural audio synthesis,''
\newblock in {\em International Conference on Machine Learning}. PMLR, 2018,
  pp. 2410--2419.

\bibitem{kong2020hifi}
Jungil Kong, Jaehyeon Kim, and Jaekyoung Bae,
\newblock ``Hifi-gan: Generative adversarial networks for efficient and high
  fidelity speech synthesis,''
\newblock {\em Advances in Neural Information Processing Systems}, vol. 33, pp.
  17022--17033, 2020.

\bibitem{wang2022campnet}
Tao Wang, Jiangyan Yi, Ruibo Fu, Jianhua Tao, and Zhengqi Wen,
\newblock ``Campnet: Context-aware mask prediction for end-to-end text-based
  speech editing,''
\newblock {\em arXiv preprint arXiv:2202.09950}, 2022.

\bibitem{tae2021editts}
Jaesung Tae, Hyeongju Kim, and Taesu Kim,
\newblock ``Editts: Score-based editing for controllable text-to-speech,''
\newblock {\em arXiv preprint arXiv:2110.02584}, 2021.

\bibitem{tan2021editspeech}
Daxin Tan, Liqun Deng, Yu~Ting Yeung, Xin Jiang, Xiao Chen, and Tan Lee,
\newblock ``Editspeech: A text based speech editing system using partial
  inference and bidirectional fusion,''
\newblock in {\em 2021 IEEE Automatic Speech Recognition and Understanding
  Workshop (ASRU)}. IEEE, 2021, pp. 626--633.

\bibitem{morrison2021context}
Max Morrison, Lucas Rencker, Zeyu Jin, Nicholas~J Bryan, Juan-Pablo Caceres,
  and Bryan Pardo,
\newblock ``Context-aware prosody correction for text-based speech editing,''
\newblock in {\em ICASSP 2021-2021 IEEE International Conference on Acoustics,
  Speech and Signal Processing (ICASSP)}. IEEE, 2021, pp. 7038--7042.

\bibitem{ren2020fastspeech}
Yi~Ren, Chenxu Hu, Xu~Tan, Tao Qin, Sheng Zhao, Zhou Zhao, and Tie-Yan Liu,
\newblock ``Fastspeech 2: Fast and high-quality end-to-end text to speech,''
\newblock {\em arXiv preprint arXiv:2006.04558}, 2020.

\bibitem{arik2018neural}
Sercan Arik, Jitong Chen, Kainan Peng, Wei Ping, and Yanqi Zhou,
\newblock ``Neural voice cloning with a few samples,''
\newblock {\em Advances in neural information processing systems}, vol. 31,
  2018.

\bibitem{chen2018sample}
Yutian Chen, Yannis Assael, Brendan Shillingford, David Budden, Scott Reed,
  Heiga Zen, Quan Wang, Luis~C Cobo, Andrew Trask, Ben Laurie, et~al.,
\newblock ``Sample efficient adaptive text-to-speech,''
\newblock {\em arXiv preprint arXiv:1809.10460}, 2018.

\bibitem{chen2021adaspeech}
Mingjian Chen, Xu~Tan, Bohan Li, Yanqing Liu, Tao Qin, Sheng Zhao, and Tie-Yan
  Liu,
\newblock ``Adaspeech: Adaptive text to speech for custom voice,''
\newblock {\em arXiv preprint arXiv:2103.00993}, 2021.

\bibitem{choi2020attentron}
Seungwoo Choi, Seungju Han, Dongyoung Kim, and Sungjoo Ha,
\newblock ``Attentron: Few-shot text-to-speech utilizing attention-based
  variable-length embedding,''
\newblock {\em arXiv preprint arXiv:2005.08484}, 2020.

\bibitem{jia2018transfer}
Ye~Jia, Yu~Zhang, Ron Weiss, Quan Wang, Jonathan Shen, Fei Ren, Patrick Nguyen,
  Ruoming Pang, Ignacio Lopez~Moreno, Yonghui Wu, et~al.,
\newblock ``Transfer learning from speaker verification to multispeaker
  text-to-speech synthesis,''
\newblock {\em Advances in neural information processing systems}, vol. 31,
  2018.

\bibitem{cooper2020zero}
Erica Cooper, Cheng-I Lai, Yusuke Yasuda, Fuming Fang, Xin Wang, Nanxin Chen,
  and Junichi Yamagishi,
\newblock ``Zero-shot multi-speaker text-to-speech with state-of-the-art neural
  speaker embeddings,''
\newblock in {\em ICASSP 2020-2020 IEEE International Conference on Acoustics,
  Speech and Signal Processing (ICASSP)}. IEEE, 2020, pp. 6184--6188.

\bibitem{skerry2018towards}
RJ~Skerry-Ryan, Eric Battenberg, Ying Xiao, Yuxuan Wang, Daisy Stanton, Joel
  Shor, Ron Weiss, Rob Clark, and Rif~A Saurous,
\newblock ``Towards end-to-end prosody transfer for expressive speech synthesis
  with tacotron,''
\newblock in {\em international conference on machine learning}. PMLR, 2018,
  pp. 4693--4702.

\bibitem{hsu2018hierarchical}
Wei-Ning Hsu, Yu~Zhang, Ron~J Weiss, Heiga Zen, Yonghui Wu, Yuxuan Wang, Yuan
  Cao, Ye~Jia, Zhifeng Chen, Jonathan Shen, et~al.,
\newblock ``Hierarchical generative modeling for controllable speech
  synthesis,''
\newblock {\em arXiv preprint arXiv:1810.07217}, 2018.

\bibitem{liu2021vara}
Peng Liu, Yuewen Cao, Songxiang Liu, Na~Hu, Guangzhi Li, Chao Weng, and Dan Su,
\newblock ``Vara-tts: Non-autoregressive text-to-speech synthesis based on very
  deep vae with residual attention,''
\newblock {\em arXiv preprint arXiv:2102.06431}, 2021.

\bibitem{zhang2019learning}
Ya-Jie Zhang, Shifeng Pan, Lei He, and Zhen-Hua Ling,
\newblock ``Learning latent representations for style control and transfer in
  end-to-end speech synthesis,''
\newblock in {\em ICASSP 2019-2019 IEEE International Conference on Acoustics,
  Speech and Signal Processing (ICASSP)}. IEEE, 2019, pp. 6945--6949.

\bibitem{kingma2013auto}
Diederik~P Kingma and Max Welling,
\newblock ``Auto-encoding variational bayes,''
\newblock {\em arXiv preprint arXiv:1312.6114}, 2013.

\end{thebibliography}

\end{document}